\documentclass{emulateapj}

\pdfoutput=1

\def\degs{\ifmmode ^{\circ}\else$^{\circ}$\fi}
\def\amin{\ifmmode ^{\prime}\else$^{\prime}$\fi}
\def\asec{\ifmmode ^{\prime\prime}\else$^{\prime\prime}$\fi}
\def\h{$^{\rm h}$}
\def\m{$^{\rm m}$}

\def\fss{\hbox{$.\!\!^{\rm s}$}}        
\def\farcs{\hbox{$.\!\!^{\prime\prime}$}}  
\newbox\grsign \setbox\grsign=\hbox{$>$}
\newdimen\grdimen \grdimen=\ht\grsign
\newbox\laxbox \newbox\gaxbox
\setbox\gaxbox=\hbox{\raise.5ex\hbox{$>$}\llap
     {\lower.5ex\hbox{$\sim$}}}\ht1=\grdimen\dp1=0pt
\setbox\laxbox=\hbox{\raise.5ex\hbox{$<$}\llap
     {\lower.5ex\hbox{$\sim$}}}\ht2=\grdimen\dp2=0pt
\def\gax{$\mathrel{\copy\gaxbox}$}
\def\lax{$\mathrel{\copy\laxbox}$}

\shortauthors{Greiner et al.}
\shorttitle{GRB 080913}

\begin{document}

\title{GRB 080913 at redshift 6.7}

\author{J. Greiner\altaffilmark{1}, T. Kr\"uhler\altaffilmark{1,2}, 
  J.P.U. Fynbo\altaffilmark{3}, A. Rossi\altaffilmark{4}, 
  R. Schwarz\altaffilmark{5}, S. Klose\altaffilmark{4},
  S. Savaglio\altaffilmark{1}, 
  N.R. Tanvir\altaffilmark{6},
  S. McBreen\altaffilmark{1}, 
  T. Totani\altaffilmark{7}, B.B. Zhang\altaffilmark{8}, 
 X.F. Wu\altaffilmark{9,10},
 D. Watson\altaffilmark{3},
 S.D. Barthelmy\altaffilmark{11},
 A.P. Beardmore\altaffilmark{6},
 P. Ferrero\altaffilmark{4},
 N. Gehrels\altaffilmark{11},
 D.A. Kann\altaffilmark{4},
 N. Kawai\altaffilmark{12},
 A. K\"{u}pc\"{u} Yolda\c{s}\altaffilmark{13},
 P. M\'esz\'aros\altaffilmark{9,14},
 B. Milvang-Jensen\altaffilmark{3},
 S.R. Oates\altaffilmark{15},
 D. Pierini\altaffilmark{1},
 P. Schady\altaffilmark{15},
 K. Toma\altaffilmark{9},
 P.M. Vreeswijk\altaffilmark{3},
 A. Yolda\c{s}\altaffilmark{1},
 B. Zhang\altaffilmark{8},
%
 P. Afonso\altaffilmark{1},
 K. Aoki\altaffilmark{17},
 D.N. Burrows\altaffilmark{9},
 C. Clemens\altaffilmark{1},  
 R. Filgas\altaffilmark{1},
 Z. Haiman\altaffilmark{18},
 D.H. Hartmann\altaffilmark{19},
 G. Hasinger\altaffilmark{1},
 J. Hjorth\altaffilmark{3},
 E. Jehin\altaffilmark{20},
 A.J. Levan\altaffilmark{21},
 E.W. Liang\altaffilmark{22},
 D. Malesani\altaffilmark{3},
 T.-S. Pyo\altaffilmark{17},
 S. Schulze\altaffilmark{4},
 G. Szokoly\altaffilmark{1,23},
 H. Terada\altaffilmark{17},
  K. Wiersema\altaffilmark{6}
 }

\altaffiltext{1}{Max-Planck-Institut f\"ur Extraterrestrische Physik, 
         Giessenbachstra\ss{}e 1, D-85740 Garching, Germany;
}

\altaffiltext{2}{Universe Cluster, Technische Universit\"{a}t M\"{u}nchen, 
   Boltzmannstra{\ss}e 2, D-85748, Garching}

\altaffiltext{3}{Dark Cosmology Centre, Niels Bohr Institute,
     University of Copenhagen, Juliane Maries Vej 30, DK-2100 K\o{}benhavn \O,
   Denmark}

\altaffiltext{4}{Th\"uringer Landessternwarte Tautenburg, Sternwarte 5, 
  D-07778 Tautenburg,  Germany}

\altaffiltext{5}{Astrophysical Institute Potsdam, D-14482 Potsdam, An der 
   Sternwarte 16, Germany}

\altaffiltext{6}{Dept. of Physics and Astronomy,
        University of Leicester, University Road, Leicester LE1 7RH, UK}

\altaffiltext{7}{Dept. of Astronomy, Kyoto University,
Sakyo-ku, Kyoto 606-8502, Japan}


\altaffiltext{8}{Dept. of Physics and Astronomy,
  Univ. of Nevada, 4505 Maryland Parkway, 
 Las Vegas, NV 89154-4002, USA}

\altaffiltext{9}{Department of Astronomy \& Astrophysics,
                             Pennsylvania State University, 525 Davey Lab,
                             University Park, PA 16802, USA}

\altaffiltext{10}{Purple Mountain Observatory, Chinese Academy of Sciences, 
  Nanjing 210008, China}

\altaffiltext{11}{NASA-GSFC, Code 661, Greenbelt, MD  20771, USA}

\altaffiltext{12}{Dept. of Physics, Tokyo Inst. of Technology,
 2-12-1 Ookayama, Meguro-ku, Tokyo 152-8551, Japan}

\altaffiltext{13}{ESO, Karl-Schwarzschild-Str. 2, 85740 Garching,Germany}

\altaffiltext{14}{Department of Physics,
        Pennsylvania State University, 525 Davey Lab,
        University Park, PA 16802, USA}

\altaffiltext{15}{Mullard Space Science Lab,  University College London,
         Holmbury St.Mary,   Dorking, Surrey, RH5 6NT, UK}

\altaffiltext{16}{Division of Theoretical Astronomy, National Astronomical 
  Observatory of Japan, 2-21-1 Osawa, Mitaka, Tokyo 181-8588, Japan}

\altaffiltext{17}{Subaru Telescope, National Astronomical Observatory
of Japan, 650 North A`oh\=ok\=u Place, Hilo, HI 96720, USA}

\altaffiltext{18}{Dept. of Astronomy, Columbia Univ., 
             1328 Pupin Physics Laboratories, New York, NY 10027, USA}

\altaffiltext{19}{Dept. of Physics and Astronomy, Clemson University,
          Clemson, SC 29634, USA}

\altaffiltext{20}{Inst. d'Astrophysique de l'Universit\'{e} de Li\`{e}ge,
                  All\'{e}e du 6 Ao\^{u}t 17, B-4000 Li\`{e}ge, Belgium}

\altaffiltext{21}{Dept. of Physics, University of Warwick, 
   Coventry CV4 7AL, UK}

\altaffiltext{22}{Dept. of Physics, Guangxi University, Guangxi 53004, China}

\altaffiltext{23}{E\"otv\"os Univ., 1117 Budapest, Pazmany P. stny. 1/A, 
  Hungary}



\begin{abstract}
We report on the detection by \emph{Swift} of GRB 080913, and subsequent
optical/near-infrared follow-up observations by GROND which led to the
discovery of its optical/NIR afterglow and the recognition of its high-z nature
via the detection of a spectral break between the $i'$ and $z'$ bands.
Spectroscopy obtained at the ESO-VLT revealed a continuum extending down to
$\lambda = 9400$ \AA, and zero flux for $7500 \AA < \lambda<9400$ \AA,
which we interpret as the onset of a Gunn-Peterson
trough at $z$=6.695$\pm$0.025 (95.5\% conf. level), 
making GRB 080913 the highest 
redshift GRB to date, and more distant than the
highest-redshift QSO.
We note that many redshift indicators which are based on promptly available
burst or afterglow properties have failed for GRB 080913.
We report on our follow-up campaign and compare
the properties of GRB 080913 with bursts at lower redshift.
In particular, since the afterglow of this burst is fainter 
than typical for GRBs, we 
show that 2\,m-class telescopes can identify most high-redshift
GRBs.

\end{abstract}



\section{Introduction}

The potential of gamma-ray bursts (GRBs) as beacons to the distant universe
has long been recognized.  The immense luminosity of both the prompt
gamma-ray emission and X-ray and optical afterglows indicates that
GRBs should, with present-day technology such as \emph{Swift}/BAT,
 be visible  out to distances of $z > 10$ \citep{lar00,gou04}.  
Due to their connection to the death of massive stars
\citep{Woosley93,galama98,hjorth03,stanek03},
long GRBs probe the evolution of cosmic star formation 
\citep{totani97,wijers98,chary07,ykb08}, 
re-ionization of the intergalactic medium
(Miralda-Escude 1998, Totani et al. 2006), 
and the metal enrichment history of the Universe 
(e.g. \citealt{hart04,fyn06,sav06,ber06,pro08}). 
Despite this potential and the high detection rate now delivered 
by \emph{Swift},
finding bursts out to the highest redshifts has proved a challenging
task.
The hitherto highest redshift burst, GRB 050904 at
z = 6.29 \citep{kka06, hnr06}, held the record for three years and
in that time only GRB 060927 at $z=5.47$ \citep{alma} came close.

\begin{table*}
\caption{Log of the observations\label{log}}
\begin{tabular}{llccc}
   \hline
   \noalign{\smallskip}
   ~~~~Day/Time  & Telescope/Instrument & Filter/  & Exposure & Brightness \\
   ~~(UT in 2008) &                   & Grism       & (s) & (mag)$^{(a)}$ \\
   \noalign{\smallskip}
   \hline
   \noalign{\smallskip}
 Sep 13 06:53--07:00 & MPI/ESO 2.2m/GROND & $g'r'i'z'$ & 4$\times$66 & $>$23.3/$>$23.6/$>$23.0/21.71$\pm$0.11
\\
 Sep 13 06:53--07:00 & MPI/ESO 2.2m/GROND & $JHK_S$ & 24$\times$10 & 
        19.96$\pm$0.03/19.64$\pm$0.05/19.17$\pm$0.11 \\
 Sep 13 07:00--07:07 & MPI/ESO 2.2m/GROND & $g'r'i'z'$ & 4$\times$66 & $>$23.3/$>$23.6/$>$22.9/22.27$\pm$0.22
\\
 Sep 13 07:00--07:07 & MPI/ESO 2.2m/GROND & $JHK_S$ & 24$\times$10 & 
         20.59$\pm$0.04/20.35$\pm$0.08/19.96$\pm$0.15 \\
 Sep 13 07:07--07:20 & MPI/ESO 2.2m/GROND & $g'r'i'z'$ & 4$\times$115 & $>$23.8/$>$24.0/$>$23.5/22.52$\pm$0.15
\\
 Sep 13 07:07--07:20 & MPI/ESO 2.2m/GROND & $JH$ & 48$\times$10 &
20.97$\pm$0.04/20.68$\pm$0.07 \\
 Sep 13 07:07--07:33 & MPI/ESO 2.2m/GROND & $K_S$ & 96$\times$10 &
20.61$\pm$0.17 \\
 Sep 13 07:20--07:33 & MPI/ESO 2.2m/GROND & $g'r'i'z'$ & 4$\times$115 & $>$23.6/$>$23.8/$>$23.3/23.20$\pm$0.26
\\
 Sep 13 07:20--07:33 & MPI/ESO 2.2m/GROND & $JH$ & 48$\times$10 &
21.58$\pm$0.11/21.14$\pm$0.07 \\
 Sep 13 08:25--08:55 & MPI/ESO 2.2m/GROND & $g'r'i'z'$ & 4$\times$375 & $>$24.4/$>$24.6/$>$24.1/24.54$\pm$0.25
\\
 Sep 13 08:25--08:55 & MPI/ESO 2.2m/GROND & $JHK_S$ & 120$\times$10  &
 22.47$\pm$0.13/22.19$\pm$0.17/21.58$\pm$0.17 \\
 Sep 13 08:55--09:40 & MPI/ESO 2.2m/GROND & $g'r'i'z'$ & 4$\times$375 + 4$\times$115& $>$24.5/$>$24.6/$>$24.2/$>$24.4\\
 Sep 13 08:55--09:40  & MPI/ESO 2.2m/GROND & $JHK_S$ & 120$\times$10 + 48$\times$10 &
 23.00$\pm$0.30/$>$22.40/$>$21.60 \\
 Sep 13 07:34--07:36 & ESO VLT/FORS2 & $z_{\rm Gunn}$ & 120 & 23.36 $\pm$ 0.13 \\
 Sep 13 07:37--07:39 & ESO VLT/FORS2 & $I$          & 80 & $>$23.8 \\
 Sep 13 07:40--07:41 & ESO VLT/FORS2 & $V$ & 80 & $>$24.0 \\
 Sep 13 07:42--07:44 & ESO VLT/FORS2 & $B$ & 150 & $>$23.6 \\
 Sep 13 07:45--07:46 & ESO VLT/FORS2 & $R$ & 60 & $>$24.1 \\
 Sep 13 08:34--08:36 & ESO VLT/FORS2 & $z_{\rm Gunn}$ & 120 & 24.31 $\pm$ 0.28 \\
 Sep 13 08:40--08:42 & ESO VLT/FORS2 & $z_{\rm Gunn}$ & 120 & 24.16 $\pm$ 0.20 \\
 Sep 13 08:52--09:22 & ESO VLT/FORS2 & grism 600z & 1800 & -- \\
 Sep 13 09:44--09:48 & ESO VLT/FORS2 & $z_{\rm Gunn}$ & 200 & 24.37 $\pm$ 0.21 \\
 Sep 13 09:52--10:02 & ESO VLT/FORS2 & grism 600z & 600 & -- \\
 Sep 13 09:02--09:12 & ESO NTT/SOFI  & $J$ & 10$\times$60 & $23.18\pm0.30$ \\
 Sep 14 08:29--09:41 & ESO VLT/FORS2 & $z_{\rm Gunn}$ & 11$\times$160 & 24.65 $\pm$ 0.37 \\
 Sep 14 12:49--15:02 & Gemini-N/NIRI & $J$ & 53 $\times$ 60 & $22.59\pm0.12$ \\
%
 Sep 15 07:12--09:27 & MPI/ESO 2.2m/GROND & $g'r'i'z'$ & 16$\times$375 & $>$24.8/$>$25.0/$>$24.3/$>$24.5 \\
 Sep 15 07:12--09:27 & MPI/ESO 2.2m/GROND & $JHK_S$ & 480$\times$10 &
   22.97$\pm$0.23/22.52$\pm$0.20/$>$21.7 \\
  Sep 15 08:45--09:51 & ESO NTT/SOFI  & $J$ & 60$\times$6$\times$10 & $22.97\pm0.16$ \\
Sep 15 12:55--14:36 & Gemini-N/NIRI & $J$ & 74 $\times$ 60 & $23.50\pm0.18$ \\
%
 Sep 16 05:25--09:15 & ESO VLT/FORS2 & grism 600z & 7$\times$1800+900 & -- \\
 Sep 16 06:38--09:46 & MPI/ESO 2.2m/GROND & $g'r'i'z'$ & 24$\times$375 & $>$25.1/$>$25.3/$>$24.8/$>$25.0 \\
 Sep 16 06:38--09:46 & MPI/ESO 2.2m/GROND & $JHK_S$ & 720$\times$10 &
   $>$23.2/$>$22.7/$>$22.0  \\
 Sep 16 12:40--14:49 & Subaru/IRCS & $J$ & 54 $\times$ 120 & 23.46 $\pm$ 0.20 \\
%
 Sep 17 06:39--07:36 & ESO VLT/HAWK-I& $J$ & 44$\times$60 & $23.48\pm0.09$ \\
 Sep 17 08:51--09:48 & ESO VLT/HAWK-I& $J$ & 44$\times$60 & $23.72\pm0.11$ \\
%
 Sep 18 07:04--08.56 & ESO VLT/FORS2 & $z_{\rm Gunn}$ &32$\times$180 & $>25.1$ \\
 Sep 18 07:59--09:00 & ESO VLT/HAWK-I & $J$ & 44$\times$60 & $24.16\pm0.13$ \\
 Sep 23 07:03--09:52 & ESO VLT/ISAAC & $J$ &64$\times$14$\times$10& $24.61\pm0.13$ \\
 Sep 29 07:19--08:03 & ESO VLT/ISAAC & $J$ &32$\times$6$\times$10   & $>$23.4 \\
 Oct 03 12:56--15:28 & Gemini-N/NIRI & $J$ & 116$\times$60 & $>$24.6 \\
  \noalign{\smallskip}
   \hline
  \noalign{\smallskip}
\end{tabular}

\noindent{
  $^{(a)}$ Not corrected for Galactic foreground reddening of E(B-V) = 0.043.
          Converted to the AB system for consistency with Fig. \ref{lc},
          using the following coefficients for the $J$ band:
          HAWK-I: +0.98 mag, Gemini-N: +0.96 mag, ISAAC: +0.96 mag, 
          GROND: +0.91 mag, SOFI: +0.96 mag,  Subaru: +0.94 mag.}
\medskip
\end{table*}

In part this owes much to the intrinsic difficulty in
locating such bursts. At $z>5.5$  afterglows become essentially
invisible to observers in the $R$-band, where much follow-up is
attempted. Often the mere time required to obtain a photometric
selection of high-z candidates is so long that 
the afterglow is too faint
for spectroscopic follow-up. Furthermore, although the
campaigns on GRB 050904 were extremely successful, this burst was
far from a typical event. Indeed, its peak optical luminosity
rivalled those of the exceptional bright GRBs 990123 and 080319B (Akerlof
et al. 1999; Kann et al. 2007a, Racusin et al. 2008).  
To identify more ``typical"
afterglows at high redshift is a much more arduous task, requiring
rapid response, multi-color observations and ultimately rapid and
deep spectroscopy. Such situations must inevitably be somewhat
fortuitous, requiring a burst to be visible to large telescopes
almost immediately after its occurrence, with good weather and
appropriate instrumentation. This perhaps explains, at least in
part, why bursts more distant than GRB 050904 have been extremely
difficult to find. Yet, the present detection rate of GRBs
at $z$ \gax 5 is about what is predicted on the basis of the 
star formation rate, so the lack of many high-z GRBs may well be 
due to the fact that there are few of them.

Here we report the discovery of  GRB 080913 with \emph{Swift} \citep{geh04},
and subsequent follow-up observations which identify this GRB to have
originated at a redshift of $z=6.7$, the highest known to date.
In Sect.~\ref{obs} we describe the observational effort in
all wavelengths from hard X-rays to near-infrared, and present
the results of these observations.
In Sect.~\ref{disc} we discuss several aspects of our observational
findings.

\section{Observations and Analysis Results}
\label{obs}

\subsection{\emph{Swift} BAT, XRT and UVOT measurements}

\emph{Swift}/BAT triggered on GRB 080913 (trigger 324561) on Sep. 13, 2008 at
T$_0$ = 06:46:54 UT \citep{schady08, sbb08}.  The BAT light curve shows multiple
overlapping peaks with a T$_{90}$ duration  (the time interval during which 90\% of
the fluence is measured) of 8$\pm$1 s.  The peak count rate was ~800 counts/s
(15-350 keV).  Using a 64 ms binned light curve, we compute spectral lags
\citep[e.g.][]{hakkila07} of 0.114$^{+0.098}_{-0.124}$ s for
the 100--150 keV vs. 50--100 keV band,
and 0.148$^{+0.094}_{-0.084}$ s for the 100--150 keV vs. 15--50 keV band, 
consistent with an independent estimate by Xu (2008).

\begin{table*}[ht]
\caption{Local photometric standards; the $JHK_S$ magnitudes are in the
  Vega system, the $g'r'i'z'$ magnitudes in AB.}
\small
\begin{tabular}{lcccccccc}
  \hline 
  \noalign{\smallskip} 
  $\!\!\!\!$No$\!\!\!\!$ & Coordinates (J2000) & $g'$ & $r'$ & $i'$ & $z'$ & $J$ & $H$ & $K_S$ \\ 
  \noalign{\smallskip} 
  \hline 
  \noalign{\smallskip}
1 & 04\h22\m56\fss1 --25\degr07\amin10\asec$\!\!$ & 
      --              &20.52$\pm$0.02&19.28$\pm$0.01 & 18.72$\pm$0.01
   & 17.50$\pm$0.04 & 16.90$\pm$0.03 & 16.90$\pm$0.06 \\
2 & 04\h22\m54\fss4 --25\degr07\amin27\asec$\!\!$ &
      --              &--              &--              &-- 
   & 20.26$\pm$0.05 & 19.31$\pm$0.06 & 18.61$\pm$0.13 \\
3 & 04\h22\m53\fss9 --25\degr07\amin37\asec$\!\!$ &
      --              &--              &--             &--
   & 19.97$\pm$0.05 & 18.91$\pm$0.05 & 18.10$\pm$0.11 \\
4 & 04\h22\m55\fss2 --25\degr07\amin39\asec$\!\!$ &
     16.97$\pm$0.01$\!$ & 16.14$\pm$0.01$\!$ & 15.82$\pm$0.01$\!$ & 15.65$\pm$0.01$\!$
   & 14.76$\pm$0.02 & 14.25$\pm$0.03 & 14.20$\pm$0.04 \\
5 & 04\h22\m55\fss3 --25\degr07\amin49\asec$\!\!$ &
     --              &--              &--             &--
   & 20.01$\pm$0.05 & 18.86$\pm$0.06 & 18.57$\pm$0.14 \\
6 & 04\h22\m54\fss2 --25\degr07\amin57\asec$\!\!$ &
   15.96$\pm$0.01$\!$ & 15.57$\pm$0.01$\!$ & 15.39$\pm$0.01$\!$ & 15.29$\pm$0.01$\!$
   & 14.54$\pm$0.02 & 14.18$\pm$0.03 & 14.19$\pm$0.04 \\
7 & 04\h22\m55\fss8 --25\degr08\amin04\asec$\!\!$ &
   18.73$\pm$0.01$\!$ & 17.26$\pm$0.01$\!$ & 16.22$\pm$0.01$\!$ & 15.76$\pm$0.01$\!$
   & 14.57$\pm$0.02 & 13.88$\pm$0.03 & 13.70$\pm$0.05 \\
8 & 04\h22\m57\fss4 --25\degr08\amin11\asec$\!\!$ &
      --              &20.90$\pm$0.04&20.20$\pm$0.03 &19.81$\pm$0.04
   & 18.34$\pm$0.05 & 17.43$\pm$0.04 & 16.71$\pm$0.07 \\
9 & 04\h22\m52\fss7 --25\degr08\amin36\asec$\!\!$ &
     --              &20.04$\pm$0.02&18.71$\pm$0.01 &18.15$\pm$0.01
   & 16.90$\pm$0.04 & 16.24$\pm$0.03 & 16.17$\pm$0.05 \\
   \noalign{\smallskip} 
   \hline
   \label{stand}
\end{tabular}
\end{table*}


The time-averaged spectrum from T$_0-3.8$ s to T$_0$+5.2 s 
can be adequately fit by a power
law with an exponential cutoff.  This fit gives a photon index of 
0.46$\pm$0.70,
and E$_{peak}$ = 93 $\pm$ 56 keV ($\chi^{2}$/dof = 38.5/56).  The total fluence
in the 15--150 keV band is (5.6 $\pm$ 0.6) $\times$ 10$^{-7}$ erg/cm$^2$, and
the 1-s peak flux measured at T$_0$+0.11 s in the 15--150 keV band is 1.4
$\pm$ 0.2 ph/cm$^2$/s.  A fit to a simple power law gives a photon index of
1.36 $\pm$ 0.15 ($\chi^2$/dof = 44.6/57)
(all the quoted errors are at the
90\% confidence level).  A combined fit of the \emph{Swift}/BAT (15--150 keV)
and Konus-Wind (20--1300 keV) data \citep{pga08} in the time interval T$_0-4.1$
to T$_0$+4.7 s using a Band function (Band et al.\ 1993; $\beta$=-2.5 fixed)
yields $\alpha$ = $-0.82_{-0.53}^{+0.75}$ and E$_{peak}$ = 121$_{-39}^{+232}$
keV (where E$_{peak}$ is the energy at which most of the power is emitted,
and $\alpha$ and $\beta$ are the low- and high-energy photon indices,
respectively), and an energy fluence in the 15--1000 keV band for the 
8.8 s interval of 9$\times$10$^{-7}$ erg/cm$^2$ \citep{pga08}.

\begin{figure*}[th]
\includegraphics[angle=270, width=18.cm]{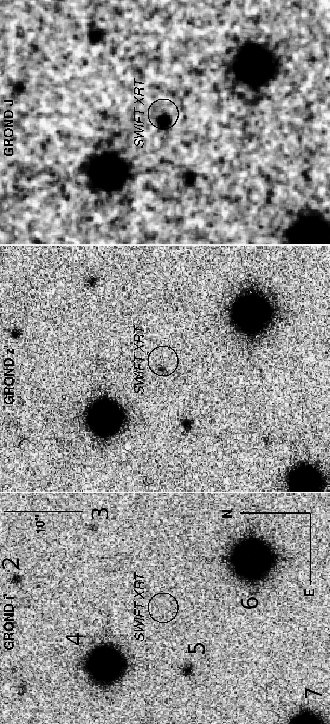}
   \caption[fc]{\small $i'$ (left), $z'$ (middle) and $J$ (right) band images 
     of the afterglow of GRB 080913 obtained with the
     7-channel imager GROND at the 2.2m
    telescope on La Silla / Chile.
    The circle denotes the {\emph Swift}/XRT error box.
    Some of the local standards of Tab. \ref{stand} are labeled 
    in the $i'$ image; the remaining ones are outside the field shown here.
    \label{fc}}
\end{figure*}

\emph{Swift} slewed immediately to the burst and the X-ray Telescope (XRT,
Burrows et al.\ 2005a) began its automated observing sequence at 06:48:30UT, 
96 s after
the trigger. A fading X-ray afterglow was detected; the UVOT-enhanced 
X-ray position  (Beardmore et al.\ 2008) is RA (J2000.0) =
04\h 22\m 54\fss66, Decl. (J2000.0) = -25\degs 07\amin 46\farcs2, with an
uncertainty of 1\farcs9 (radius, 90\% confidence; see Fig. \ref{fc}).

The X-ray light curve (Fig. \ref{lc}) shows an 
initial decay rate ($F \propto t^{-\alpha}$)
of $\alpha \sim  1.2$ from $\sim$100 s to $\sim$300s and 
from $\sim$400 to $\sim$1100 s, with a likely 
small flare with the same decay slope 
afterwards. At around T$_0$+1.8 ks ($\Delta T /
T \sim 0.3$), there is a substantial flare with 
a factor of $\sim 5$ increase in count rate.
The evolution thereafter is difficult to characterize due to the sparse
coverage: it could be a continued decay from 2 ks to $\sim$100 ks at the same 
slope of 1.2 but with an offset suggestive of energy injection, followed 
by a plateau, or it can be described as decaying at a slope of about 
1.2 (Beardmore \& Schady 2008) 
from 1200 s to 10$^6$ s with two flares super-imposed.

The X-ray spectrum (using the \emph{Swift} software package v2.9 and its
associated set of calibration files) from ${\rm T_0}+108$ s to 1920 s is well
fit by an absorbed power law of photon index $\Gamma = 1.66\pm0.14$.  A separate
fit of the early flare and preflare times gives the same photon index
(1.69$\pm$0.25).  The column density is consistent with the Galactic value of
$3.2\times 10^{20}\thinspace {\rm cm^{-2}}$ in the direction of the burst
(Kalberla et al.\ 2005). The observed $0.3-10.0$ keV flux at
this time was $8.0^{+1.1}_{-1.0} \times 10^{-12} \thinspace {\rm
erg}\thinspace {\rm cm^{-2}}\thinspace {\rm s^{-1}}$. This corresponds to an
unabsorbed  flux of $8.5^{+1.1}_{-1.0} \times
10^{-12} \thinspace {\rm erg}\thinspace {\rm cm^{-2}}\thinspace {\rm s^{-1}}$
($0.3-10.0$ keV).

UVOT observations in white-light (100 s exposure) started 105 s after the 
BAT trigger, and subsequently all UVOT filters were used. The afterglow
was not detected in any of the UVOT filters and the 3 $\sigma$
limiting white magnitude for the first finding chart exposure is
 $>$20.92; see Oates \& Schady (2008) 
for the detailed upper limits in each filter.

\subsection{Optical/NIR photometry}

\subsubsection{Observations}

GROND, a simultaneous 7-channel imager \citep{gbc08} mounted at  the 2.2\,m
MPI/ESO telescope at La Silla (Chile), started observing the field at 06:52:57
UT, about 6 min after the GRB.  The imaging sequence began with 66 s
integrations in the $g'r'i'z'$ channels with gaps of about 16--18  s due to
detector read-out and preset to a new telescope dither position.  After about
15 min, the exposure time was increased to 115 s, and after another 22 min to
375 s.  In parallel, the three near-infrared channels $JHK_S$ were operated with
10 s integrations, separated by 5 s due to read-out, data-transfer and 
$K_S$-band
mirror dithering.  At 09:46 UT, the start of nautical twilight (Sun is 12$\degs$
below horizon), GROND switched to the ``NIR-only'' mode, in which the CCDs of
the 4 visual channels are switched off, and only imaging in $JHK_S$ is 
performed.
Observations finally stopped completely at 10:07 UT.  Further GROND imaging was
performed on Sep. 15, 07:12--09:28 UT, and Sep. 16, 06:38--09:47 UT.

Imaging was also secured immediately after the burst and on the subsequent
nights at the VLT (ESO) with the Focal Reducer and Spectrograph FORS in filters
$BVRIz$, and with ISAAC and HAWK-I in $J$.  
Further imaging was obtained with NTT/SOFI,
Subaru/IRCS, and Gemini-N/NIRI in the $J$ band (Tab. \ref{log}).

Data reduction was done in a standard way using IRAF routines.
Photometric calibration of the GROND $g',r',i',z'$ bands was performed 
using the Sloan
spectrophotometric standard star SA95-142, which was observed shortly after
the GRB 080913 field. Magnitudes are therefore given in AB magnitudes  since
this is the natural photometric system for GROND (see Greiner et al.
2008a). 
In order to match the different $z'J$-band filters used,
the field calibration of the GROND filter magnitudes was compared 
against the VLT (or Gemini/Subaru) magnitudes using $\sim$50 field stars.
For instance, the rms scatter between the GROND $z'$ and FORS2 Gunn-$z$ 
was $<0.06$ mag, so no color transformations were applied.
Calibration of the field in $JHK$ was performed using 2MASS stars which
were chosen as close as possible to the GRB afterglow
(and with a nicely sampled point-spread function), to reduce the error in 
the calibration due to flatfielding, which is 
(for GROND) 0.5--1.0\% on small
scales and around 2\% over the whole array.
The magnitudes of the selected 2MASS stars were then transformed 
into the GROND filter system and finally into
AB magnitudes using $J$(AB) = $J$(Vega) + 0.91, $H$(AB) = $H$(Vega) + 1.38,
$K$(AB) = $K$(Vega) + 1.81 (for details, see Greiner et al. 2008a). In this
way, a set of nine secondary standard stars was created
(Tab.~\ref{stand}).

\subsubsection{The photometric redshift estimate}

After stacking the GROND exposures taken during 
the first 16 min (06:53--07:07 UT), the GROND pipeline reduction
\citep{kyk08} found a faint source 
at RA (J2000.0) = 04\h 22\m 54\fss74 Decl.
(J2000.0) = --25\degs 07\amin46\farcs2 (0\farcs5 error), which was only
detected in the $z'JHK_S$-bands, but not in shorter-wavelength bands 
(Figs. \ref{fc}, \ref{sed}), and thus suggested a
redshift above 6 (Rossi et al.\ 2008; Greiner et al.\ 2008b). This was confirmed
by VLT/FORS2  $RzIVBR$ imaging, starting 45 minutes after the burst, 
the only detection being in the $z$ band with z(AB) = 23.1 \citep{vfm08}. 
A fit to the simultaneously obtained 7-filter GROND spectral 
energy distribution (SED), using Hyper-z (Bolzonella et al.\
2000) results in a photometric redshift of z = 6.44$\pm$0.30 (Fig. \ref{sed};
Greiner et al.\ 2008b).

\subsubsection{The afterglow light curve}

\begin{figure}
\centering
\includegraphics[width=12.cm]{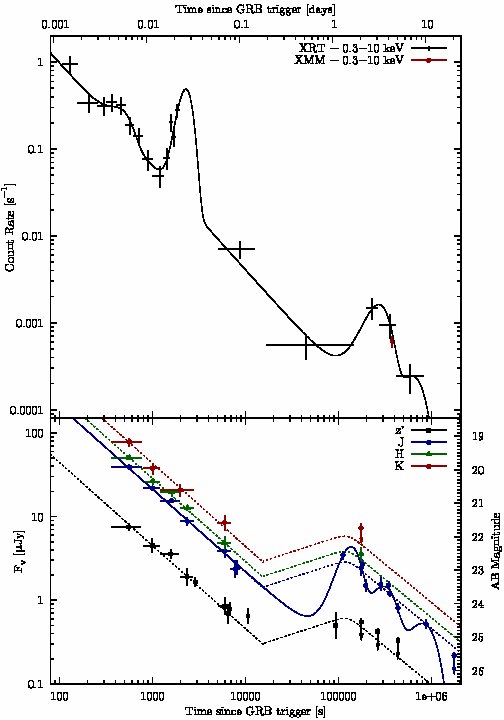}
    \caption[lc]{Optical/NIR (bottom) and X-ray light curve 
    (top) of the GRB 080913 afterglow. 
     The XMM data point has been converted to the Swift/XRT count rate
     axis using the best-fit spectral model.
    Several deviations from a canonical power law decay are apparent
    in both, X-rays as well as optical. The early X-ray light curve
    ($T_0$+200 s until $T_0$+2000 s) is modeled by two log-Gaussian flares.
    For the optical light curve two different fits are shown: 
     one for the plateau interpretation (dotted lines) 
     and one for the flares interpretation (three log-Gaussians; solid line).
    The first model is motivated by
    the scenario of a plateau and late energy injection. In the second model, 
    which assumes that X-ray flares are associated with optical flares,
    the last NIR data point
    is too bright, and consequently needs a third flare. The late Gemini
    upper limit supports this interpretation.
    The sum of the three log-Gaussian profiles roughly accounts for
    the flare at X-rays at the same time interval. 
    \label{lc}}
\end{figure}

\begin{figure}[hb]
\centering
\includegraphics[angle=270, width=.60\columnwidth]{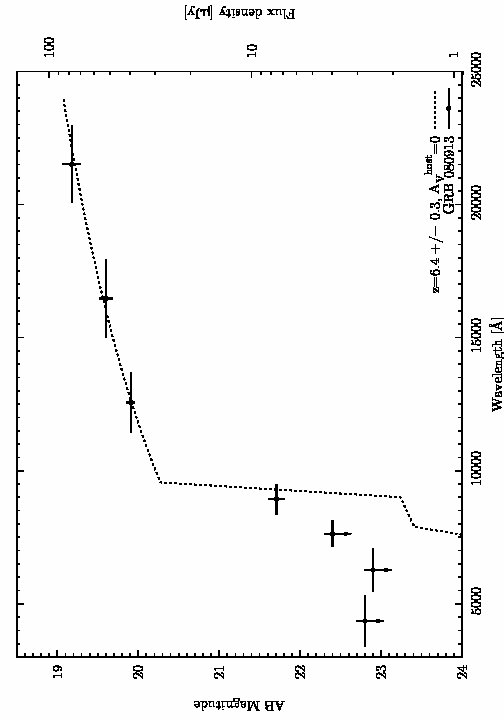}
    \caption[sed]{Spectral energy distribution of the afterglow
    of GRB 080913, obtained about 30 min after the GRB.
    The drop-out shortward of the $z'$-band is clearly visible.
    This information was used to trigger spectroscopy with VLT/FORS2
    which then started with grism 600z, selected as compromise between
    spectral resolution and faintness of the afterglow, 
    about 2 hrs after the GRB. Shifting the later VLT magnitudes
    in $BVRIz$ to the earlier time of the GROND measurement,
    the upper limits would be $B > 22.5$, $V > 22.6$, $R > 22.5$ and 
    $I > 22.0$, thus not more constraining than the GROND limits.
         }
    \label{sed}
\end{figure}

The optical/NIR light curve (Fig. \ref{lc}) 
during the first few hours is described
by a power law of 0.98$\pm$0.05. The third GROND data point
at 2000 s post-burst is simultaneous to the X-ray flare, and is clearly
above that power law decay, suggesting that we see enhanced optical
emission from that X-ray flare (Fig. \ref{lc}). 
If we ignore this third data point,
the reduced $\chi^2$ improves substantially, and the new power law decay is
1.03$\pm$0.02. After the GRB location became visible again after 2 days
(clouds prevented observations on Sep. 14), at $\sim$180 ks post burst, 
the optical/NIR brightness as measured by
GROND is nearly identical to that at 10 ks, suggesting a prolonged plateau 
phase (Fig. \ref{lc}). Later monitoring with GROND, VLT, NTT, Gemini-N and 
Subaru shows pronounced variability which can be, in principle, fit by
three log-Gaussian flares. At the same time, also the X-ray light curve shows
an enhanced flux level. It is likely, though it cannot be proven, that
the X-ray light curve also consisted of flares, which are not resolved
due to the faintness and the somewhat poorer
sampling as compared to the optical/NIR band.
In any case, the enhanced X-ray flux lets us prefer that the
optical lightcurve is probably a flare rather than a plateau.
Our $J$ band data point ($J = 24.61\pm0.13$) at $8\times10^5$ s
is again well above a power law extrapolation
after the two flares.

\subsubsection{The GROND-derived afterglow broad-band spectrum}

We fit the NIR SED (the $z^\prime$ and Gunn-$z$ bands
are affected by Ly-$\alpha$ and are not included)
with no extinction and with three different 
dust models (LMC, MW, SMC). Here we assume that the afterglow
spectrum should have a power law shape, with $F \propto \nu^{-\beta}$
(note that the spectral index $\beta$ is related to the photon index
$\Gamma$ from the X-ray spectral fitting by $\beta$ = $\Gamma$ - 1). 
The best-fit power law has $\beta$ = 1.12$\pm$0.16.
No evidence for extinction in the host frame is found.

\subsection{Optical spectroscopy}

\begin{figure}[hb]
 \centering
\includegraphics[width=15.cm]{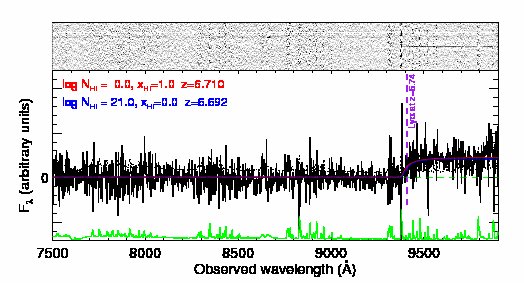}
    \caption[ospec]{Optical spectrum of the afterglow of GRB 080913
     obtained with VLT/FORS2. The spectrum shown corresponds to 
     the 2400 s exposure on Sep 13,   2008.
    The break due to the Ly$\alpha$ forest is clearly visible at 9400 \AA. 
    The bottom green curve shows sky spectrum, and the dotted line
    superimposed on the spectrum shows the error spectrum (noise) 
    after the sky subtraction.
    Two model fits are shown (see text).
    \label{redsh}}
\end{figure}

Immediately after determining the photo-z, we triggered our
target-of-opportunity program at ESO (PI: J. Greiner), and obtained an 
optical spectrum of the
afterglow of GRB 080913 in the 7500--10500 \AA\ (grism 600z) 
region with FORS2/VLT on Sep.
13, 2008. Only one 1800 s and one 600 s exposures were possible before dawn.
The spectrum revealed a continuum disappearing bluewards of a break around 9400
\AA. Interpreting this break as the onset of the Lyman-$\alpha$ forest a
preliminary redshift estimate of $z=6.7$ was inferred \citep{fgk08}.  A second
set of exposures (7$\times$1800 s) was acquired on Sep 16, 2008, with
identical settings.  The first night's spectrum is shown in Fig. \ref{redsh}.
The afterglow is also detected in the spectrum from 2 nights later, but
with smaller signal-to-noise ratio.

In order to constrain the redshift we first tried to search the spectrum for a
significant metal absorption line. Unfortunately, no such line seems to be
present in the spectrum. Therefore, the redshift can be based on the shape of
the break at 9400 \AA\ only: the flux is zero in the interval 7500--9400 \AA, 
and clearly non-zero above 9400 \AA, strongly supporting the detection of 
the Lyman break, i.e. a high-z object. 
We first tried a simple cross-correlation analysis
using the ``fxcor'' 
task in IRAF. As templates we used the spectra of GRBs 050730
(z = 3.969) and 060206 (z = 4.048, redwards of the center of the DLAs 
of these spectra). This
analysis leads to a redshift of $z=6.69\pm0.02$. 

In order to make a more refined
analysis we follow \citet{tkk06} and fit the spectrum using two different
assumptions about the origin of the break. The damping wing can either be
interpreted as that of a Damped Lyman-$\alpha$ (DLA) absorber in the host galaxy
\citep{jakobsson06} or that of IGM neutral hydrogen \citep{madau}. We performed
a chi-square analysis with the three parameters of $N_{\rm HI}$ (column density
of DLA), $x_{\rm HI}$ (neutral fraction of IGM hydrogen), and 
$z$ (redshift of the GRB). We assume that the DLA has the same
redshift $z$, and the neutral IGM hydrogen is distributed from
$z_{IGM, l}$ to $z$, meaning that there is no ionized bubble around the 
host galaxy.
We assume $z_{IGM, l} = 6.0$, but this has little influence on the fit if
     $z - z_{IGM, l} > \sim 0.3$ (Totani et al. 2006).
No prior is assumed for these model parameters.
  We obtained the best-fit at 
$(\log N_{\rm HI}/{\rm cm}^{-2}, x_{\rm HI}, z) = (19.74, 1.0, 6.71)$, 
in which the damping 
wing is dominated by IGM. 
On the other hand,
a DLA dominating case of (21.41, 0.001, 6.68) is also consistent with the data
(1.2 $\sigma$ deviation from the best fit).  Marginalizing $N_{\rm HI}$ and
$x_{\rm HI}$, we obtain the 95.4\% confidence limits on $z$ as $6.67 < z <
6.72$, where the DLA and IGM are dominant in the lower and upper limits,
respectively.  

When we fix $x_{\rm HI} = 0.001$ (almost fully ionized IGM with
negligible effect on the damping wing), we obtain 95.4\% limits on $N_{\rm HI}$
as $20.29 < \log N_{\rm HI}/{\rm cm}^{-2} < 21.41$
marginalizing the redshift, with the best fit value
of $\log N_{\rm HI}/{\rm cm}^{-2} = 20.99$ at $z = 6.69$.  
When we assume that $N_{\rm HI}$ is
negligibly small, we obtain a 95.4\% lower-limit of $x_{\rm HI} > 0.35$ again
marginalizing redshift.  This means that, if the column density of the DLA in
the host galaxy is sufficiently small to have a negligible effect on the 
observed damping wing, we need a significantly neutral IGM indicating that the
re-ionization has not yet been completed at $z=6.7$.  Although this possibility
is completely dependent on the assumption about DLA column density, it is not
in contradiction with the 95\% upper bound of $x_{\rm HI} < 0.60$ at $z = 6.3$
obtained from GRB 050904 (Totani et al.\ 2006). 
A significantly high neutral fraction of IGM at $z \gtrsim 6.7$ has also
been implied from evidence for the GP damping wing in the spectra of 
z=6 quasars (Mesinger \& Haiman 2004, 2007)
and the luminosity function evolution of Ly-$\alpha$ emitters (Kobayashi et al.
2007; Ota et al. 2008).
 These claims remain controversial, since different sightlines to quasars can
produce large
 variations in the Ly-$\alpha$ absorption spectra (Bolton \& Haehnelt 2007), 
and the apparent
evolution of the Ly-$\alpha$ luminosity function may be explained by the 
evolution of the mean IGM density alone (Dijkstra et al. 2007).

\subsection{The X-ray to optical SED}

\begin{figure}
   \centering
  \includegraphics[width=0.45\textwidth,angle=270]{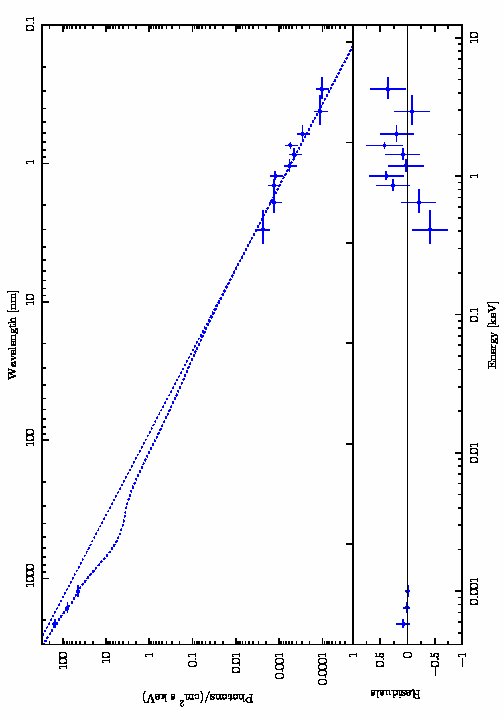}
   \caption{Combined \emph{Swift}/XRT and GROND spectral energy distribution for
      the time interval 200-1800 s after the burst.  Shown is the
       fit with a single power law (for parameters see text).
    \label{SED}}
\end{figure}

The combined XRT and GROND/VLT spectral energy distribution
was fit (except the $z'$ band since it is affected by Ly-$\alpha$)
in three separate time intervals:
the early decay except the second flare at T$_0$+2000 s, i.e. the
interval 360--1200 s, 
the intermediate brightness level (5.16--12.6 ks),
and at very late times (16.7--550 ks).
We used a single as well as a broken power law, modified by dust
extinction in the rest-frame FUV (zdust in XSPEC) for Milky Way, SMC and LMC,
and hydrogen absorption at X-rays. The hydrogen absorption was fit 
for all the data, and then fixed at that value for the individual
intervals. 
For the single power law fit, we obtain a photon index of 1.7$\pm$0.1
and a marginal reddening E(B-V) of 0.096$\pm$0.045 ($\chi^{2}$/dof = 8.4/10)
assuming LMC dust in the GRB host (see Fig. \ref{SED}).
The results for SMC or MW dust are the same within the errors.
Forcing the dust extinction to be zero results in a somewhat worse fit
($\chi^{2}$/dof = 15/11).
We note that the GROND-XRT SED for this first time interval 
likely is affected by enhanced X-ray emission from the early flare 
(at 300--700 s), thus resulting in a flatter spectral slope
than derived from the GROND data alone.

Using a broken power law fit instead 
and fixing the XRT power law slope
to 1.7 (see above), we obtain a photon index of 2.1$^{+0.45}_{-0.03}$
for the GROND data (corresponding to the energy index of 1.1; see above) 
and negligible dust, i.e.  E(B-V) =  0.026$^{+0.027}_{-0.054}$.
This implies that the early X-ray afterglow emission is contaminated by 
the flare emission, leading to a flatter X-ray-to-optical SED.
The goodness of the fit is not better than the single power law fit
($\chi^{2}$/dof = 11/9), so a broken power law is anyway not required by the 
data. For the two late-time SEDs, the uncertainties are larger due to the
lower signal-to-noise ratio, and thus 
are equally well fit by a single power law.

\subsection{Late-time XMM-Newton observation}

A triggered XMM-{\it Newton} observation was performed at 380 ks
post-burst. The afterglow is clearly detected. 
There is no evidence for any variability in the XMM lightcurve.
The spectrum is acceptably fit with a power-law with fixed Galactic
absorption (3.2$\times$10$^{20}$ cm$^{-2}$). The 
best-fit photon index is $\Gamma$ = 2.0$\pm$0.2.
The unabsorbed flux is 2.2$\times$10$^{-14}$ erg/cm$^2$/s  in the 
0.3--10.0 keV band.

\section{Discussion}
\label{disc}

With $z$=6.695$\pm$0.025, GRB 080913 is the most distant burst detected so far
(and actually the second most distant spectroscopically confirmed object
after a galaxy at $z=6.96$,
as well as the most distant X-ray source) for which a redshift has been
determined \citep{iye08}.
In a concordance cosmology model (Spergel et al. 2003), 
with H$_0$ = 71 km/s/Mpc, $\Omega_M$ = 0.27,
$\Omega_{\Lambda}$ = 0.73, the corresponding luminosity distance is 67 Gpc
($2\times10^{29}$ cm), and
the age of the Universe at that time is 825 Myr (6\% of the present age).

\subsection{Comparison to other GRBs}

\subsubsection{GRB properties}

With the above numbers,
the isotropic energy release is E$_{iso} \approx 7\times 10^{52}$ erg 
\citep[1 keV $-$ 10 MeV; see also][]{pga08}.
This is typical of the
long GRB population at large.  Compared to the properties of GRB 050904, the
other previous GRB at z$>$6, GRB 080913 has a substantially shorter duration,
lower gamma-ray luminosity as well as much dimmer ($\sim$5 mag at early times)
afterglow.  Thus, it appears that gamma-ray bursts at this early epoch show the
same large diversity as the low-$z$ bursts.

\subsubsection{On the burst duration}

The observed bimodal hardness-duration distribution of the prompt 
emission from GRBs has led to the distinction between 
long- (\gax2 s) and short-duration (\lax2 s) bursts
\citep{mgi81,norr84,kou93}.
 Long-duration gamma-ray bursts are thought to arise in jets
created by the collapse of a massive star, short-duration bursts have been
suggested to emerge from a compact binary merger 
\citep{Eichler,Paczynski87,Woosley93}.  GRB 080913
is, in its rest frame, relatively short and hard, with a rest frame duration
around 1 s, and rest frame $E_{peak}$ $\sim 990$ keV for a Band function 
fit (Pal'shin et
al. 2008). Other moderately high redshift bursts with rest-frame duration
shorter than 2 s are 
GRBs 060206 (T$_{90}$ = $7\pm2$ s, z = 4.0), 
051016B (T$_{90}$ = $4\pm0.1$ s, z = 0.94) , 
050406 (T$_{90}$ = $5\pm1$ s, z = 2.44), 
050416A (T$_{90}$ = $2.4\pm0.2$ s, z = 0.6535), 
000301C (T$_{90}$ $\sim$ 2 s, z = 2.03), 
040924 (T$_{90} \sim$ 1.5 s, z = 0.859) and
050922C (T$_{90}$ = 5$\pm$1 s, z = 2.2)
(Levan et al.\ 2007), as well as
GRBs 
 060223: (T$_{90}$ = 11$\pm$2 s, z = 4.41),
 060926: (T$_{90}$ = 8.0$\pm$0.1 s, z = 3.20),
 070506: (T$_{90}$ = 4.3$\pm$0.3 s, z = 2.31),
 071020: (T$_{90}$ = 4.2$\pm$0.2 s, z = 2.145),
 080520: (T$_{90}$ = 2.8$\pm$0.7 s, z = 1.545).
These bursts all show several of the typical signs of a
massive stellar origin, e.g., star-forming host galaxies, a gas-rich
line-of-sight, and (in the case of GRB 050416A) an associated supernova. 
Since the long- to short-duration separation at $\sim$2 s was derived
in the observer's frame for BATSE bursts for which redshifts are not
available in general, we
consider it most plausible that GRB 080913, despite its intrinsically relatively
short duration, is a member of the long-duration population of GRBs.  
Also, GRB 080913 is compatible with the previously known lag-luminosity 
correlation of long-duration bursts (,
and is also consistent with the Amati
relation. While a merger origin for GRB 080913 can not be definitely
ruled out, the short duration also leads to an interesting question for the 
collapsar scenario: How can a massive star produce a burst as short as 
1 second?

\subsubsection{Afterglow properties}

\begin{figure}
   \centering
  \includegraphics[width=0.55\textwidth,angle=0]{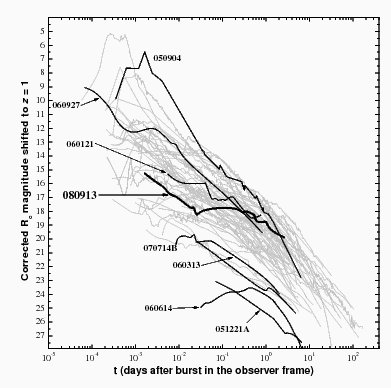}
   \caption{The optical/NIR afterglow of GRB 080913 (thick black line), 
    shifted to  the $R_C$
   band and to a redshift $z=1$, and compared with other GRB afterglows,
   also at $z=1$ \citep{Kann2007,Kann2008}. At early times, the optical/NIR
   afterglow of GRB 080913 is fainter than the bulk of the long GRBs
   with detected afterglows, and at late times it is brighter.
   With the possible exception of GRB 060121 \citep[if it lies at
   $z=4.6$,][]{dup06}, the optical/NIR afterglow of GRB 080913 is also much
   brighter than even those of bright short GRBs.
\label{olccomp}}
\end{figure}

In order to compare the afterglow of GRB 080913 with those of previous GRBs 
we analyze its intrinsic properties both in the optical
as well as the X-ray regime within large afterglow samples obtained for other
GRBs.
Using $\beta=1.1$ we
shift the light curve to $z=1$ following \cite{Kann2006}. The
result is presented in Fig. \ref{olccomp}, and compared with other
 GRB afterglow light curves. At early
times, the afterglow is fainter than the mean of the sample, and especially
much fainter than the optical afterglows of the GRBs with the second and third
highest redshifts, GRB 050904 and GRB 060927, respectively.
At one day after the GRB at $z=1$, following a strong
rebrightening, $R_C=18.80\pm0.13$
and $M_B=-24.1\pm0.2$, which is brighter than the mean of the sample of
\cite{Kann2007}, $\overline{M_B}=-23.0\pm0.4$, but not exceedingly so. Indeed,
even after the strong rebrightening, there are several afterglows that are
considerably brighter at this time in the same frame.

\begin{figure}
 \centering
  \includegraphics[width=0.55\textwidth,angle=0]{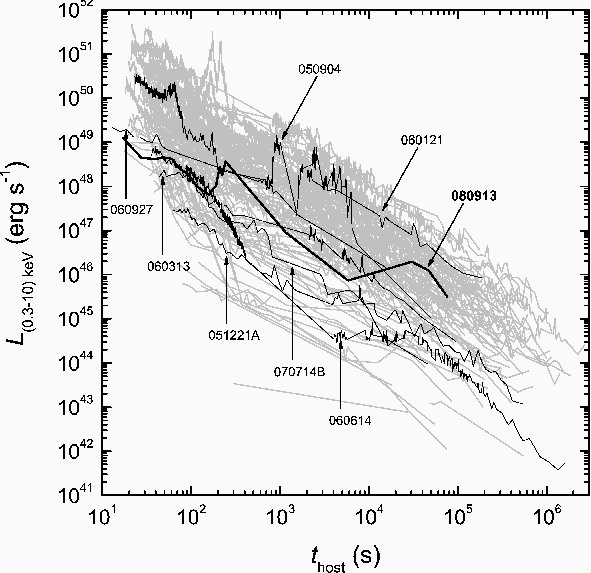}
\caption{The X-ray afterglow of GRB 080913 in comparison with 110 X-ray
afterglows with known redshift discovered by \emph{Swift} since 2005.
\label{XrayAGs}}
\end{figure}

Fig. \ref{XrayAGs} shows the luminosity evolution of the X-ray afterglow of
GRB 080913 in comparison to the population of 110 X-ray afterglows of
\emph{Swift} bursts with known redshift detected by July 2008. The original
data were obtained from the X-ray light curve repository (Evans et al. 2007).
Luminosities were calculated following Nousek et al. (2006). The
spectral slope $\beta$ of the SED  was obtained by fitting an absorbed
power-law, $F (E_{\rm obs} ) \sim \exp(- N_{\rm H}^{\rm Gal} \,\sigma(E_{\rm
obs})) \exp(- N_{\rm H}^{\rm host} \,\sigma(E_{\rm host})) \, \nu^{-\beta}$
with $E_{\rm host} = E_{\rm obs}\,(1+z)$,  while fixing the Galactic column density
to the value given by Kalberla et al. (2005). In
Fig.~\ref{XrayAGs} the low luminosity end is mainly occupied by short burst
afterglows, which on average separate clearly from their long burst
cousins. Some short bursts are indicated (GRBs 051221A, 060313, 060121 at
$z$=4.6: de Ugarte Postigo et al. 2006, or GRB 070714B) as well as the  
second and third most distant 
bursts GRB 050904 (Haislip et al. 2006; Kawai et al. 2006) and GRB 060927
(Ruiz-Velasco et al. 2007).  These events, together with GRB 080913, indicate a
rather broad luminosity distribution even at high
redshift. Fig.~\ref{XrayAGs} makes clear that similar to the optical/NIR
bands, the rest-frame X-ray afterglow of GRB 080913 is rather faint. 
Compared to the long
burst ensemble it belongs to the low luminosity members, while compared to the
short burst ensemble it occupies the high-luminosity region. 

\subsection{Modelling of the afterglow}

The temporal and spectral power law indices of the early ($t<t_{b1}\sim10^4$ s)
optical/NIR afterglow, $\alpha_{\rm{opt},1}=1.03\pm0.02$ and
$\beta_{\rm{opt}}=1.12\pm 0.16$, are consistent with the closure relation
$\alpha_{\rm{opt},1}-1.5\beta_{\rm{opt}}\sim-0.5$, which
corresponds to the standard model with the optical band above the
cooling frequency $\nu_c$ and the typical synchrotron frequency
$\nu_m$. The electron energy distribution index $p\sim2.2$,
inferred from $p=(4\alpha_{\rm{opt},1}+2)/3$ and/or
$p=2\beta_{\rm{opt}}$ of the standard model prediction
(M\'{e}sz\'{a}ros \& Rees 1997; Sari et al. 1998), is identical to
the canonical one. 

The early optical/NIR data can also constrain the
values of other parameters (e.g., initial isotropic kinetic energy
$E_i$, energy fraction in electrons $\epsilon_e$ and that in
magnetic fields $\epsilon_B$) of the forward shock model. 
If the circum-burst
environment is interstellar medium, the requirement of
$\nu_{\rm{opt}}
> {\rm{max}}(\nu_m, \nu_c)$ at $t = 560$ s and the $H$ band flux density
$F_{\nu_H} \sim 4$ $\mu$Jy at $t = t_{b1}$, can be translated to a
set of loose constraints. In general, $10^{-5} < \epsilon_B <
\epsilon_e < 1$, $E_{i,53} \leq 1.0$ and/or a density of $n\geq 100$ cm$^{-3}$
can fit the afterglow, which is quite reasonable except maybe
for the relatively high density. If the
circum-burst environment is a free wind, then the same
observational constraints would require a more ad hoc set of
parameters, i.e. $10^{-5} < \epsilon_B < \epsilon_e < 1$,
$E_{i,53} \leq 1.0$ and $A_{\ast} \geq 1.0$
(where $A_{\ast} = A / (5 \times 10^{11} {\rm g \thinspace cm^{-1}})$, and
$A$ is the normalization in the density profile $\rho = A r^{-2}$). 
While a high-density
medium is expected at high-$z$ (e.g. Gou et al. 2007), a large
wind parameter $A_{\ast} \geq 1.0$ 
is not expected in view of the
low stellar metallicity at high-$z$. We therefore conclude that
the afterglow modeling favors a constant density medium, although
a stellar wind medium is not ruled out.

After the early decay, the afterglow intensity enters a plateau/flares
phase. Due to the lower signal-to-noise ratio the optical to X-ray SED
with a photon index of 1.94$\pm$0.20 now has larger error bars, but is still
consistent with $\beta_{OX} \sim 1.0$
which indicates that the emission of both bands possibly has the same origin.

For the flares interpretation (see solid line in Fig. \ref{lc}),
the late time optical/NIR light curve is fit with  log-Gaussian
functions superimposed to the initial unbroken power law decay
with temporal index fixed to be $\alpha_{\rm{opt},1}=1.03$. The
best-fit amplitude and mid-time of the optical flares are
$F_{\nu_J,p} = 2.9 \pm 1.2$ $\mu$Jy, $t_{p} = (1.33 \pm 0.17) \times
10^{5}$ s for the first flare,  $F_{\nu_J,p} = (0.95 \pm 0.28)$
$\mu$Jy, $t_{p} = (3.00 \pm 0.43) \times 10^{5}$ s for the second flare,
and $F_{\nu_J,p} = (0.52 \pm 0.05)$
$\mu$Jy, $t_{p} = (7.7 \pm 0.4) \times 10^{5}$ s for the third flare. 
The late time X-ray flare happens almost simultaneously
with the optical flares. These late flares together with the early
two X-ray flares (at T$_0$ + 300 s and T$_0$ + 2000 s)
are very likely the emission from late internal
shocks, as already detected in many previous \emph{Swift} GRBs,
long or short, and in GRB 050904 at comparable redshift (Burrows
et al. 2005b; Barthelmy et al. 2005;  Watson et al. 2006;
Cusumano et al. 2006).
The central engine of GRB 080913 in this case has accelerated and
emitted relativistic ejecta intermittently for a whole period of
at least $\sim4\times10^4$ s in its rest frame.

For the plateau interpretation (see dotted lines of Fig. 2), the
late time optical light curve can be fit by a broken power law
with $\alpha_{\rm{opt},2} = -0.19^{+0.15}_{-0.16}$ during the
plateau until $t = t_{b2} = 1.16^{+0.19}_{-0.17}\times10^{5}$ s, and
$\alpha_{3} = 0.92^{+0.09}_{-0.08}$ after the plateau. A joint fit
combining the optical data with the X-ray data does not change the above
values significantly, though the rising in X-rays between
T$_0$ + 50 ks to T$_0$ + 200 ks is more pronounced than at optical/NIR 
wavelengths. The nearly achromatic beginning and ending
times of the plateau, no significant spectral evolution across the
break, and nearly same decay rate before and after the plateau 
(if averaging over the short-scale variability)
all suggest that the re-brightening is due to the forward shock
emission with a continuous energy injection, either due to a long
term central engine injecting energy in the form of Poynting flux
(Dai \& Lu 1998; Zhang \& M\'{e}sz\'{a}ros 2002) or due to spread
of the ejecta Lorentz factor distribution (Rees \&
M\'{e}sz\'{a}ros 1998; Sari \& M\'{e}sz\'{a}ros 2000; Zhang \&
M\'{e}sz\'{a}ros 2002). In the former model, the Poynting flux
luminosity from a spinning down central magnetar or black hole is
 $L = L_0(1 + t/T_0)^{-2}$, while in the latter model,
the mass-Lorentz factor distribution may be modeled as $M(>
\Gamma) \propto \Gamma^{-s}$. The total energy in the forward
shock during the plateau phase increases with time as $t^{a}$.
Using the optical spectral index, we can derive
$a=2\Delta\alpha_{\rm{opt}}/(1+\beta_{\rm{opt}})=1.15\pm0.16$,
where
$\Delta\alpha_{\rm{opt}}=\alpha_{\rm{opt},1}-\alpha_{\rm{opt},2}=1.22\pm0.15$.
Therefore, the total energy has increased by a factor of
$(t_{b2}/t_{b1})^a\sim12$ compared to its initial value $E_i$.
According to Table 2 of Zhang et al. (2006), we obtain
$q=1-a=-0.15\pm0.16$, $s=(7a+3)/(3-a)=6.0\pm0.8$ (ISM) and
$s=(3a+1)/(1-a)=-29.7\pm31.8$ (wind). The value $q\sim0$ is quite
consistent with the Poynting flux injection model for times
earlier than the central engine spin down time scale ($t<T_0$).

\subsection{On high-redshift indicators}

Various redshift indicators have been proposed in the past with the aim to
select, based on high-energy data, candidates for high-redshift GRBs.  Those
include \emph{Swift}/BAT properties like burst duration (T$_{90} > 60$ s), 
peak photon flux ($<$ 1 ph/s/cm$^2$), and spectral steepness 
(Campana et al. 2007; Salvaterra et al.
2007; Ukwatta et al. 2008), 
 slowly rising GRB profiles (BAT image triggers as
opposed to rate triggers), or the ``pseudo''-redshift (in this case
z=2.5--6.0) based on the peak energy  (Pelangeon 2006). 
For GRB 080913, most
of these indicators have failed (in fact also for GRB 060927 at z=5.47),
except for the excess-N$_{\rm H}$ method (Grupe et al. 2007).
As these criteria are statistical in
nature, outliers are possible. However, the parameters of this burst
demonstrate that one should not bias the search for high-$z$ events on those
criteria.  High-z GRBs are so rare that we cannot afford to loose the rare
events that do occur. The trigger criteria we are setting up for our
follow-up observations should rather include false
triggers than exclude the true ones. The only secure way to find these high-$z$
events is by simultaneous optical/NIR follow-up observations to establish their
drop-out nature from the colors of the afterglows.

\begin{table}
\caption{Gamma-ray bursts at $z > 5$.}
\begin{tabular}{lcl}
  \hline 
  \noalign{\smallskip}
  GRB   & redshift & Reference \\
  \noalign{\smallskip} 
  \hline 
  \noalign{\smallskip}
 060522 & 5.11$\pm$0.01 & \cite{cenko06} \\
 050814 & 5.77$\pm$0.12 & \cite{jlf06,cur08} \\ 
 060927 & 5.467        & \cite{alma} \\
 050904 & 6.295$\pm$0.002 & \cite{kka06} \\
 080913 & 6.695$\pm$0.025 & this work \\
   \noalign{\smallskip} 
   \hline
\end{tabular}
\label{grbz5}
\medskip
\end{table}

\subsection{GRBs at $z>6$}

GRB 080913 is a proof that GRBs do occur at $z\gtrsim6.7$ and hence
that a mechanism for star formation
and evolution leading to a burst of this duration was in place at this time.
As we
move to higher and higher redshifts the fraction of massive stars that are of
population III (low metallicity) will increase. 
When this happens is strongly dependent on the 
strength of winds from the first stars and the efficiency of mixing of the 
metals \citep{Scannapieco}. In most models population III stars and stars
only enriched by population III stars can be found down to $z\gtrsim5$.
Simulations of supernovae from population III stars suggest that the
exploding stars will have hydrogen and helium rich outer 
layers \citep{ohkubo,lawlor}
and hence that they probably will not make GRBs at least by the collapsar
mechanism. Concerning GRBs from binary population III stars it is debated how
efficient they are in producing GRBs \citep{bl06,belczynski}.

Detecting high-redshift GRBs with \emph{Swift} and measuring their redshifts
with ground-based spectroscopy is of substantial interest because of the link
between long-duration GRBs and the star formation rate density (SFRD).  
Indeed, the properties of GRB hosts
and the distribution of GRB metallicities are consistent with the assumption
that GRBs trace the bulk of the star-formation at $z\gtrsim3$
\citep{jakobsson05,fynbo08}.  
The existence of now 5 GRBs at $z>5$ (Tab. \ref{grbz5}), among them two 
at $z>6$ out of a total of $\sim$150 GRBs with redshift estimates suggests that
the global SFRD at $z>5$ declines slowly and has a
substantial value. 
Empirical calibrations of the GRB rate to the SFR
at different redshifts have been attempted, and suggest that the high-$z$ SFR
must be high to accommodate the several observed high-$z$ GRBs 
(Chary et al. 2007, Y\"uksel et al.\ 2008). 
Our detection of a GRB at $z$=6.7 strengthens this result.
This is also similar to the SFRD in the model assumed by Bromm \&
Loeb (2006) that predicts 
$\sim 10\%$ of the \emph{Swift} GRBs are at $z>5$.

The discovery of bursts like GRB 080913 also allows to study, in an unbiased
way, some of the main questions in the evolution of the Universe: where did most
of the star formation happen at $z>6$, and what was the nature of the sources
responsible for the re-ionization? There is evidence that the bright $z>6$
galaxies discovered using color-color (drop-out) selection or more advanced
photometric redshifts are too rare to provide the total star formation rate as
well as to \emph{have done} the re-ionization (e.g., Bouwens et al.\ 2007). GRB
measurements provide the tool to find the more typical galaxies responsible for
the bulk of the production of ionizing photons \citep{alma}, and will allow
further study of these galaxies in the future (e.g. with JWST or the
$\sim$30\,m ground-based telescopes).

The re-ionization history of the Universe is currently not well
constrained by observations. The WMAP data (Spergel et al. 2007)
and the SDSS quasar data (Fan et al. 2006),
and the GRB 050904 constraint (Totani et al. 2006)
can accommodate several distinct
re-ionization scenarios (e.g. Holder et al. 2003). In order to
identify the correct scenario, bright beacons in the dark era (e.g.
$z=7-1100$) are needed. The discovery of GRB 080913 above the highest
$z$ for QSOs (CFHQS J2329-0301 at $z=6.427\pm 0.002$; Willot et al. 2007)
 re-enforced the possibility to use high-z GRBs to uncover
the cosmic re-ionization history.  
The detection of lensed star formation at $z > 7$ (e.g., Bouwens et
al. 2008) suggests that star formation in fact
took place as early as $z \sim 8-10$ (at $t < 0.63$ Gyr).
With hints for the first stars 
having formed as early as 20\lax\ z \lax 60 (Kogut et al. 2003,
Bromm \& Loeb 2006, Naoz \& Bromberg 2007), 
GRBs are believed to exist as early
as $z\sim 15-20$, and their gamma-ray and IR emissions are bright
enough to be detected by the current instruments (Lamb \& Reichart
2000; Ciardi \& Loeb 2000; Gou et al. 2004). A caveat has been that
the GRB host DLAs may have a large column density that would bury the
signature of IGM absorption, as in the case of GRB 050904 (Totani et
al. 2006). Cosmological SPH simulations (Nagamine et al. 2008), on
the other hand, suggest that the GRB host DLA columns should degrade
at high-z, a prediction dictated by the fundamental structure
formation theory. The low N(HI) associated with GRB 080913 ($5\times 10^{20}
{\rm cm^{-2}}$ for the best fit value and $1.4 \times 10^{21}{\rm cm^
{-2}}$ for the maximum value) is generally consistent with such a
prediction, though the medium N(HI) for $z>4$ GRBs  (10$^{21.45}$ cm$^{-2}$) 
is consistent with the medium value (10$^{21.3}$ cm$^{-2}$)
for the total GRB population ($2<z<6.3$).
In any case, high-z GRBs offer the promising possibility to
probe cosmic re-ionization. Brighter GRBs at higher $z$'s with a
negligible N(HI) would allow mapping the IGM ionization fraction as a
function of $z$ in the dark era, and hence, greatly constrain the
possible scenarios of cosmic re-ionization.

\section{Conclusions}

We have presented the observations that led to the discovery of
GRB 080913 and the recognition of it being at redshift 6.7,
the most distant GRB so far. This discovery was possible first of
all due to the excellent properties of the \emph{Swift} mission
with its high sensitivity for GRBs and excellent localization
capabilities based on the X-ray afterglows alone. Furthermore,
this study demonstrates that 2\,m class 
telescopes are sufficient to properly identify and determine
reliable photometric redshifts of the most distant GRB afterglows.
Here it is noteworthy that the afterglow of GRB 080913 was several
magnitudes fainter than that of GRB 050904, the only other 
$z>6$ GRB discovered so far. Hence, we can probe a significant 
fraction of the luminosity function of afterglows at these
redshifts with 2\,m class telescopes.
We also note that many previously proposed high-z ``indicators''
primarily based on properties of the prompt emission
have failed for GRB 080913, leaving photo-z determinations
the only reliable method available.

While the afterglow of this GRB
was particularly faint, 
we note that
metallicity measurements as well as the determination of
the neutral hydrogen fraction should be possible in general
(as for GRB 050904)
though likely not in all cases. It is important to recognize, however,
that studying the  neutral hydrogen fraction in the highest redshift
quasars is hopeless, and for galaxies complicated by their faintness
and more structured continuum shape \citep{mlz08}.
Planned instrumentation in the near future with higher 
efficiency and higher spectral resolution 
 such as X-shooter \citep{Kaper08} will help substantially
in exploiting the promise that GRBs have as probes of the re-ionization epoch. 



\acknowledgements 

We are very grateful for the excellent support by the La Silla and Paranal
Observatory staff, in particular to Stephane Brillant, Michelle Doherty, 
Carla Gil, Rachel Gilmour, Swetlana Hubrig, Heidi Korhonen, Chris Lidman, 
Emanuela Pompei, Julia Scharw\"achter,  and  Linda Schmidtobreick. 
We acknowledge discussions with J.P. Norris, T. Sakamoto,
D. Grupe, E. Rol, N. Masetti and E. Palazzi. 
TK acknowledges support by
the DFG cluster of excellence 'Origin and Structure of the Universe',
APB from STFC, and.  
PMV from the EU under a Marie Curie Intra-European 
Fellowship, contract  MEIF-CT-2006-041363.
Part of the funding for GROND (both hardware as well as personnel)
was generously granted from the Leibniz-Prize (DFG grant HA 1850/28-1)
to Prof. G. Hasinger (MPE).
This work is partly based on observations collected at the European Southern
Observatory, Chile under ESO proposal Nos. 081.A-0135, 081.A-0856 and
081.A-0966.  The Dark Cosmology Centre is funded by the DNRF.

\bigskip

{\it Facilities:} \facility{Max Planck:2.2m}, \facility{Swift}.




\end{document}